%% file: main.tex
\documentclass[sigconf]{acmart}

\usepackage{enumerate}
\usepackage{amsmath}
\usepackage{multirow, makecell}
\usepackage{pifont}
\usepackage{subfig}
\usepackage{graphicx}
\usepackage{colortbl}
\usepackage{balance}
\usepackage{xcolor}
\usepackage[space]{grffile}
\usepackage{lscape}
\usepackage{caption}
\usepackage{commath}
\usepackage{tablefootnote}
\usepackage{tikz}
\usepackage{wrapfig}
\usepackage{parallel}

\AtBeginDocument{%
  }

\newcommand{\changed}[1]{\textcolor{black}{#1}}

\newcommand{\ourmethod}{\textit{DALTON}}

\copyrightyear{2024}
\acmYear{2024}
\setcopyright{rightsretained}
\acmConference[MOBILEHCI Adjunct '24]{26th International Conference on Mobile Human-Computer Interaction}{September 30-October 3, 2024}{Melbourne, VIC, Australia}
\acmBooktitle{26th International Conference on Mobile Human-Computer Interaction (MOBILEHCI Adjunct '24), September 30-October 3, 2024, Melbourne, VIC, Australia}
\acmDOI{10.1145/3640471.3680243}
\acmISBN{979-8-4007-0506-9/24/09}

\begin{document}

\title{Exploiting Air Quality Monitors to Perform Indoor Surveillance: Academic Setting}

\author{Prasenjit Karmakar}
\email{prasenjitkarmakar52282@gmail.com}
\affiliation{%
	\institution{IIT Kharagpur}
	\country{India}
}

\author{Swadhin Pradhan}
\email{swadhinjeet88@gmail.com}
\affiliation{%
	\institution{Cisco Systems}
	\country{USA}
}

\author{Sandip Chakraborty}
\email{sandipc@cse.iitkgp.ac.in}
\affiliation{%
	\institution{IIT Kharagpur}
	\country{India}
}

\input{Sections/abstract}


\maketitle

\input{Sections/intro}

\input{Sections/apparatus}

\input{Sections/pilot}
\input{Sections/dataset}
\input{Sections/eval}
\input{Sections/conclusion}
\input{Sections/ethics}
\input{Sections/acknow}

\bibliographystyle{ACM-Reference-Format}
\bibliography{reference}
\end{document}

%% file: Sections/abstract.tex
\begin{abstract}
Changing public perceptions and government regulations have led to the widespread use of low-cost air quality monitors in modern indoor spaces. Typically, these monitors detect air pollutants to augment the end user's understanding of her indoor environment. Studies have shown that having access to one's air quality context reinforces the user's urge to take necessary actions to improve the air over time. Thus, user's activities significantly influence the indoor air quality. Such correlation can be exploited to get hold of sensitive indoor activities from the side-channel air quality fluctuations. This study explores the odds of identifying eight indoor activities (i.e., enter, exit, fan on, fan off, AC on, AC off, gathering, eating) in a research lab with an in-house low-cost air quality monitoring platform named \ourmethod{}. Our extensive data collection and analysis over three months shows 97.7\% classification accuracy in our dataset.
\end{abstract}

%% file: Sections/intro.tex
 \section{Introduction}
\label{sec:intro}

Air quality Monitors are becoming ubiquitous in modern indoor spaces due to government regulations and growing awareness among the general population. In 2023, this market was estimated to be US\$ 5006 million, which is expected to expand up to US\$ 11672 million within the next decade~\cite{indoorairmarket}. A typical air monitoring solution~\cite{airknight,airthings,pranaair_sensi_plus} provides the end user with an understanding of their pollution exposure. Such devices send data to cloud servers for storage and to offload computational overheads of analyzing long-term data rather than doing it on-device to maintain a low-power and portable form factor. Cloud storage and computing enable the development of online dashboards and mobile applications to visualize the overall pollution patterns~\cite{brazauskas2021data}, trigger alerts and notifications~\cite{zhong2020hilo}, derive countermeasures~\cite{zhu2022dynamic}, etc., reinforcing the end user towards improving air quality for healthier indoors. However, sharing such indoor pollution data with a third party can be concerning due to the high correlation between the performed indoor activities and changes in pollution signature~\cite{verma2021racer,fang2016airsense}. Therefore, the data can be used as a side-channel to eavesdrop on indoor activities and carry out surveillance without the user's consent.


In the last decade, several studies have explored activity monitoring, and the literature can be grouped into two categories based on the utilized modality: (i) Direct video and audio approaches, (ii) Side-channel approaches like wearables, millimeter-wave (mmWave) radars, radio frequency identification (RFID) tags, air quality monitors, etc. Direct video~\cite{khan2024human} or audio~\cite{10.1145/3242587.3242609} based approaches are privacy-intrusive and usually require explicit permission from the end user to be operational. Users prefer to avoid capturing video or audio data in private spaces. 

\changed{In contrast, wearables (e.g., smartwatches, smart glasses, earbuds, etc.) and radio frequency (RF) based sensing (i.e., with new generation cellular technology~\cite{fang2022joint}) devices are becoming ubiquitous in our day-to-day lives. Several studies~\cite{yuan2024self,10.1145/3645091} have reported the effectiveness of wearables in monitoring an individual's daily activities. However, due to limited battery life, the invasive nature of wearables, and discomfort for long-term usage renders such devices unsuitable for continuous monitoring. On the other hand, relatively new RF-based methods are pervasive and can monitor multiple individuals in an area. Nowadays, such systems are extensively being used in industrial or warehouse applications to track packages~\cite{10.1145/3411822}, monitor workers, etc. Studies have also explored mmWave in typical indoors to identify user activities~\cite{10577316} and perform remote patient monitoring~\cite{10.1145/3556558.3558578} in smart homes scenarios. With the rollout of 6G cellular networks, RF-based monitoring systems are yet to be widely adopted in general consumer space.}

\begin{figure*}
        \centering
	\captionsetup[subfigure]{}
		\subfloat[Experimental setup\label{fig:setup}]{
			\includegraphics[width=0.19\textwidth,keepaspectratio]{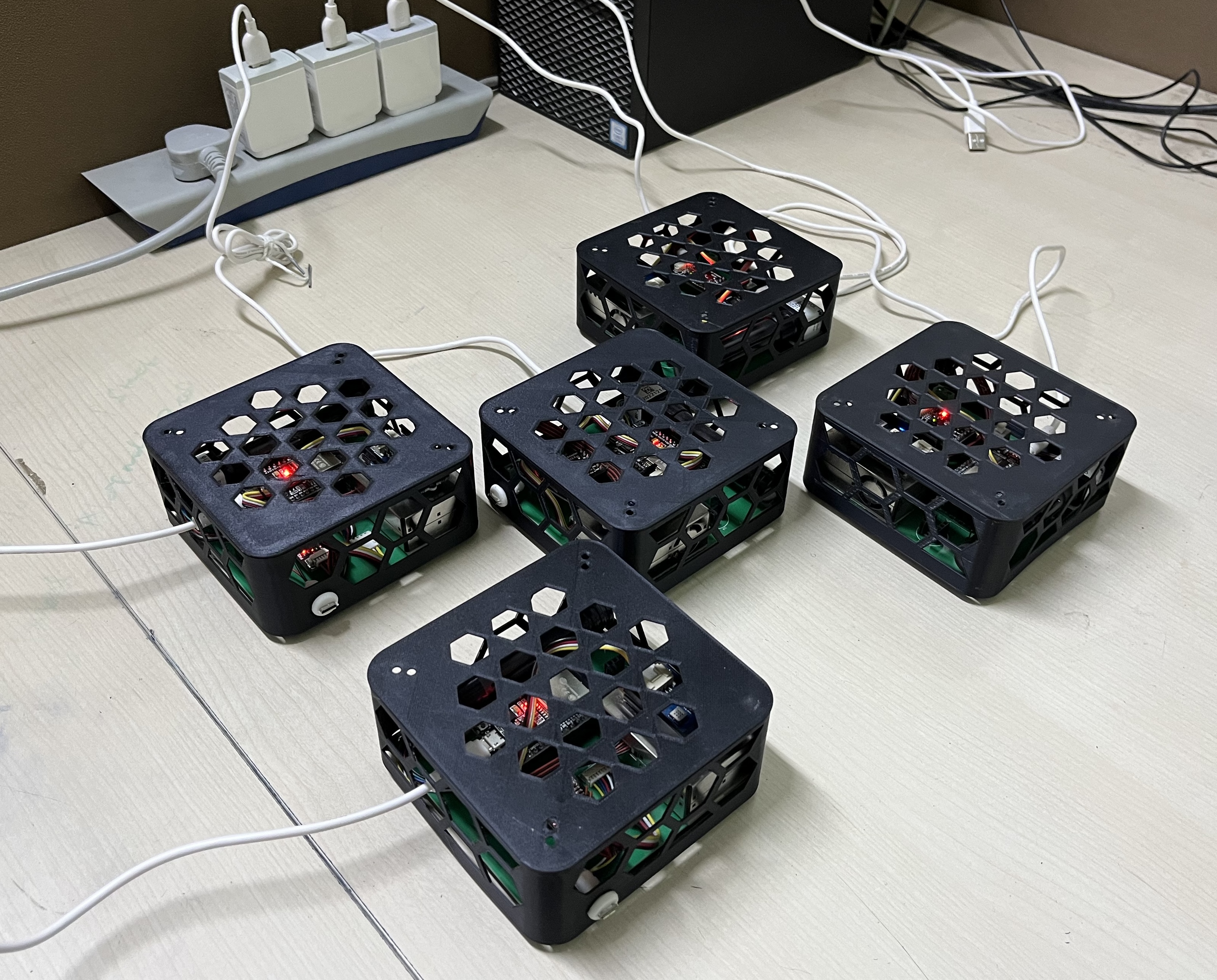}
		}
		\subfloat[CO\textsubscript{2}\label{fig:co_reading}]{
			\includegraphics[width=0.19\textwidth,keepaspectratio]{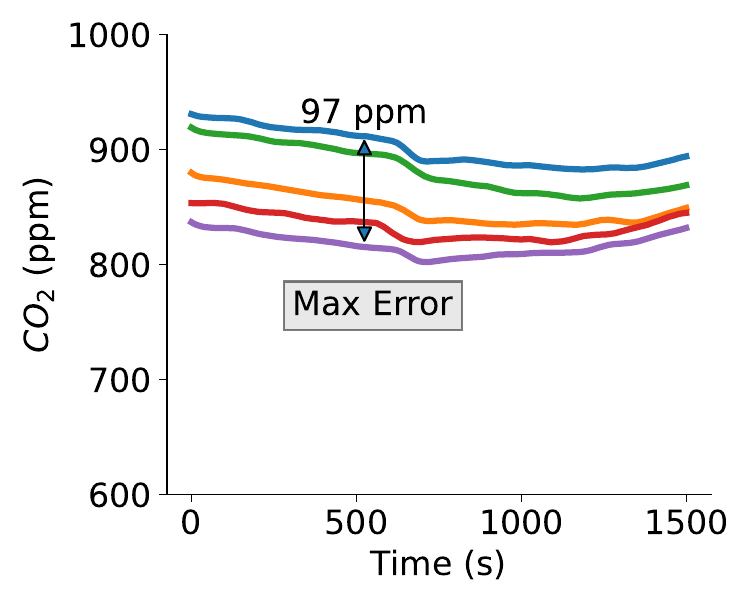}
		}
            \subfloat[PM\textsubscript{2.5}\label{fig:co_reading}]{
			\includegraphics[width=0.19\textwidth,keepaspectratio]{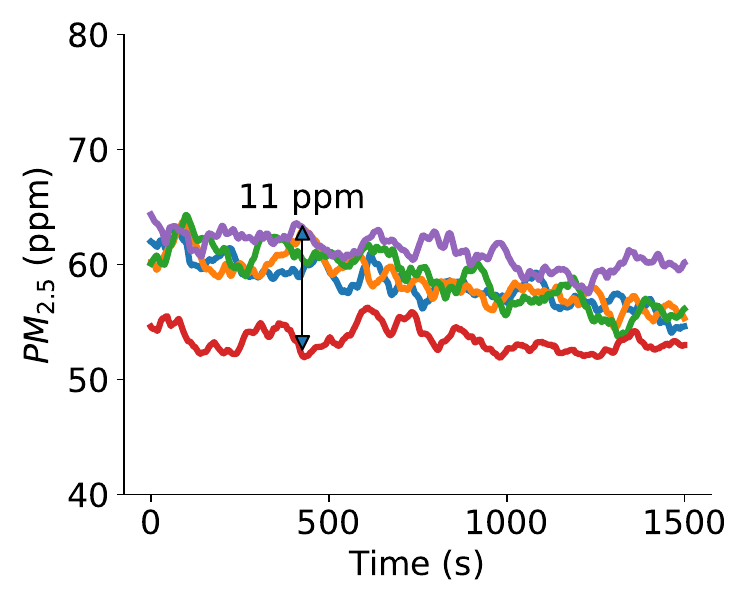}
		}
            \subfloat[Temperature\label{fig:co_reading}]{
			\includegraphics[width=0.19\textwidth,keepaspectratio]{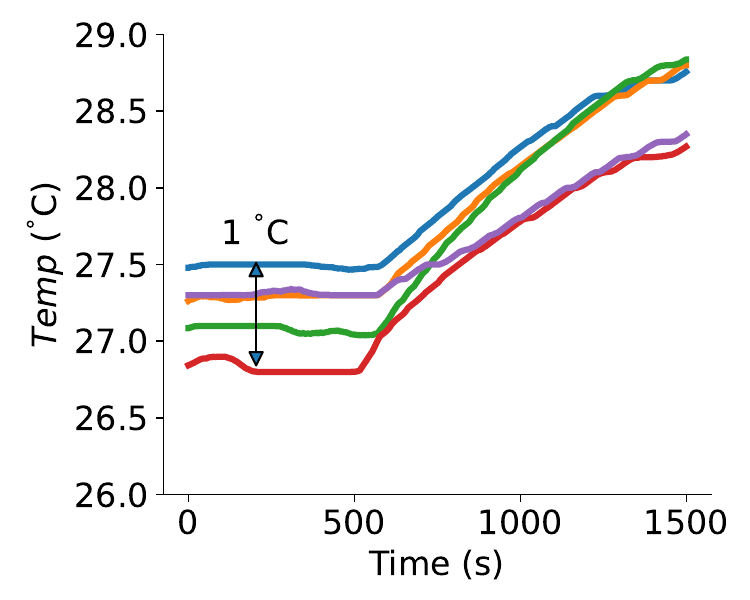}
		}
            \subfloat[VOC\label{fig:co_reading}]{
			\includegraphics[width=0.19\textwidth,keepaspectratio]{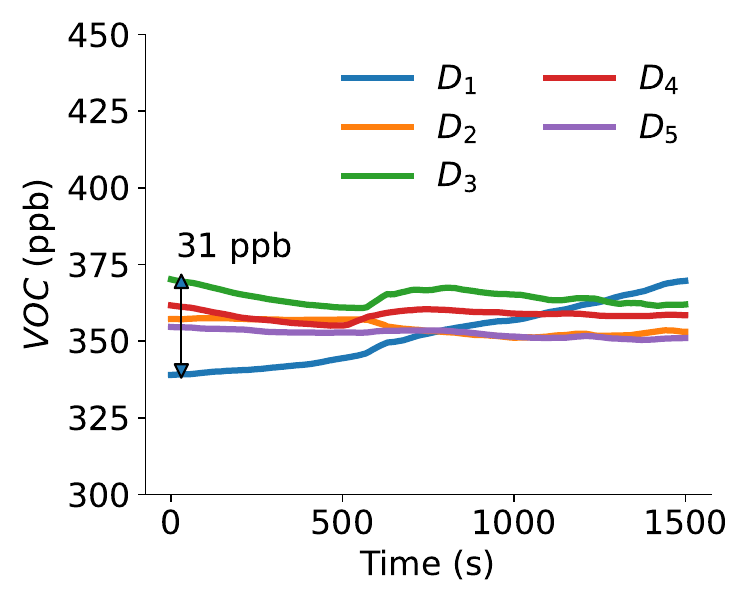}
		}
	\caption{Readings from five colocated \ourmethod{} devices indicate the variability across sensors made by the same vendor. The maximum error between two devices is within the error margin, as reported by the vendor.}
	\label{fig:colocate}
\end{figure*}

\changed{Unlike wearables or RF-based sensing, air quality monitors have started to combine with household and office appliances (e.g., air conditioners, air purifiers, kitchen chimneys, etc.) and are becoming an integral part of modern indoor spaces. Apart from measuring pollutants to improve indoor air, few studies~\cite{gambi2020adl,fang2016airsense,verma2021racer} have explored correlations between users' activities and fluctuations in air pollutants. For instance, \cite{fang2016airsense} have detected cooking, smoking, and spraying activities. \cite{verma2021racer} has identified indoor meetings, walking in the corridor, cooking, window open, etc. Detecting side-channel activity from air quality data preserves the identity of the user, but it can be utilized in restricted environments (e.g., Research Lab) or private ones (e.g., Household) to collect sensitive information. A third party can further exploit such information to perform targeted advertisement or long-term user behavior modeling~\cite{canalytica}. However, very limited studies in the literature evaluate such side-channel applications of air quality data for activity monitoring.}

\changed{Aiming to address this research gap, we propose to use air quality modalities to monitor specific lab activities through indoor surveillance. We have deployed an in-house air monitoring platform named \ourmethod{}~\cite{karmakar2024exploring} in four corners of the research lab to capture the overall fluctuations in indoor pollutants (i.e., carbon dioxide, volatile organic compounds, particulate matter) and environmental parameters (i.e., temperature, humidity) that get influenced by the in lab activities. We have considered eight indoor activities that can be grouped into three categories based on surveillance motive: (i) Engagement and occupancy (i.e., enter, exit), (ii) Occupant behavior (i.e., fan on/off, AC on/off), (iii) Prohibited practices (i.e., gathering, eating). Seven researchers voluntarily participated in the study and annotated their activities over a three-month data collection period. In our pilot study, we establish the influence of activities on the indoor environment. Further, our extensive evaluation with simple off-the-shelf machine learning models shows a maximum accuracy of 97.7\% in classifying the considered activities, indicating the efficacy of air quality data for indoor monitoring.}

%% file: Sections/apparatus.tex
\section{Experimental Setup}
\changed{The \ourmethod{} air quality monitoring platform is developed as a part of our prior work~\cite{karmakar2024exploring}. The platform is highly extensible and plug-\&-play, enabling us to effectively monitor pollution dynamics of indoor spaces with a multi-module deployment. These \ourmethod{} sensing modules are portable and easy to deploy, having the footprint of a typical lunchbox (112 mm $\times$ 112 mm $\times$ 55 mm).} The module is equipped with multiple research-grade sensors that together measure the concentration of pollutants, such as \textit{Particulate matter} (PM\textsubscript{x}), \textit{Nitrogen dioxide} (NO\textsubscript{2}), \textit{Ethanol} (C\textsubscript{2}H\textsubscript{5}OH), \textit{Vola\textit{tile organic compounds} (VOCs), \textit{Carbon monoxide} (CO), }Carbon dioxide (CO\textsubscript{2}), etc., with \textit{Temperature} (T) and \textit{Relative humidity} (RH). We utilize the ESP-WROOM-32 chip as the on-device processing unit that packs a dual-core Xtensa 32-bit LX6 MCU with WiFi $2.4$GHz HT40 capabilities. The connectivity board is a two-layer printed circuit board (FR4 material). The outer shell of the module is a 3D printed (PLA+ material) hollow structure with honeycomb holes so that the air within the module is the same as outside, resulting in unbiased measurement of pollutants at a sampling frequency of $1$Hz.\changed{The module draws a maximum power of 3.55 watts from a standard wall socket to be operational and relies on the wifi network to communicate the pollutant readings to the data logging server.}

Although the sensors are factory-calibrated, we have explicitly calibrated each sensor to ensure the correctness of the measurements. We have calibrated the PM\textsubscript{2.5}, Temperature, and Relative Humidity sensors using a reference Airthings device~\cite{airthings}. For the CO\textsubscript{2} readings, we have calibrated the MH-Z16 sensor to Zero point (400 ppm) and SPAN point (2000 ppm) as an initial step before the deployment. Further, we have turned on the self-calibration mode of the sensor so that it can judge the zero point intelligently and do the calibration automatically every 24 hours. The other measurements, such as NO\textsubscript{2}, C\textsubscript{2}H\textsubscript{5}OH, VOC, and CO are one-point calibrated before deployment and periodically cross-checked with a calibrated \ourmethod{} device during the data collection period. \figurename~\ref{fig:colocate} shows measurements from five colocated modules, validating acceptable variability across sensors made by the same vendor.

%% file: Sections/pilot.tex
\begin{figure*}
        \centering
	\captionsetup[subfigure]{}
		\subfloat[CO\textsubscript{2} during exam\label{fig:exam}]{
			\includegraphics[width=0.33\textwidth,keepaspectratio]{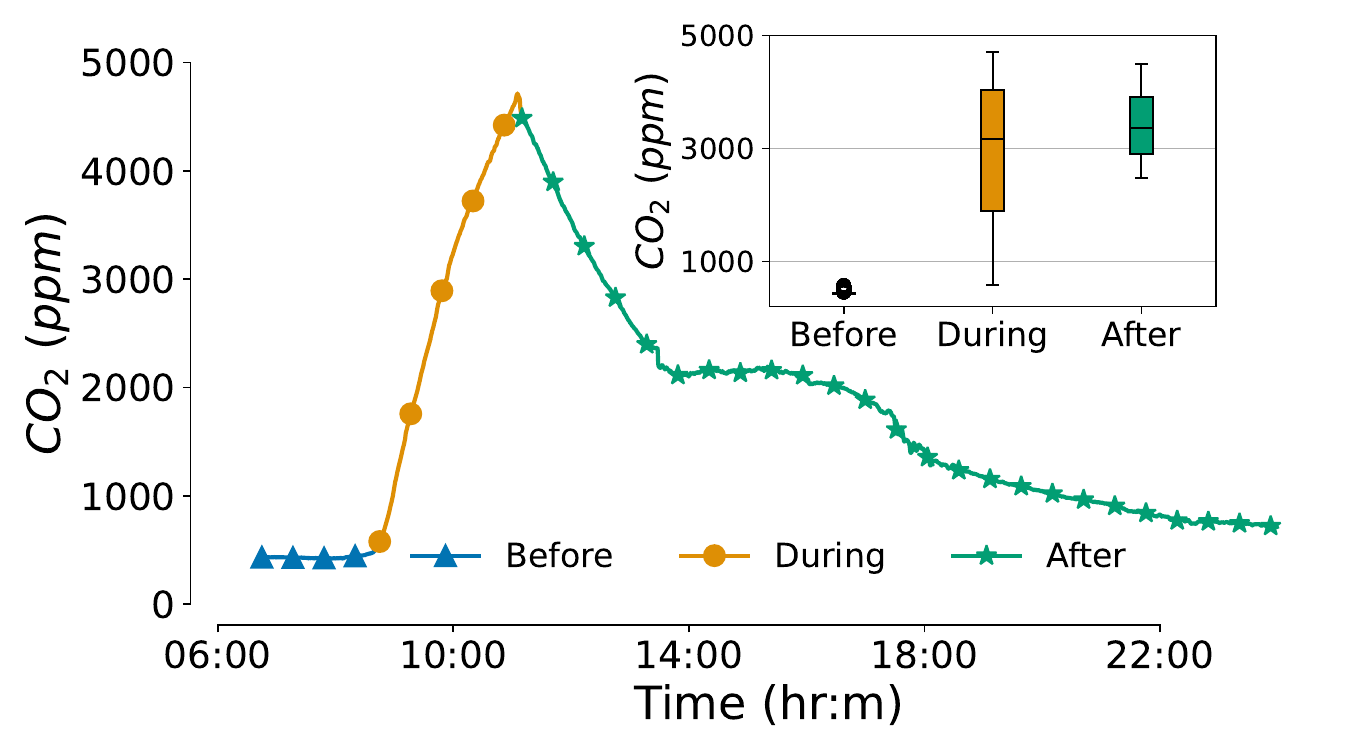}
		}
		\subfloat[AC on/off\label{fig:ac}]{
			\includegraphics[width=0.25\textwidth,keepaspectratio]{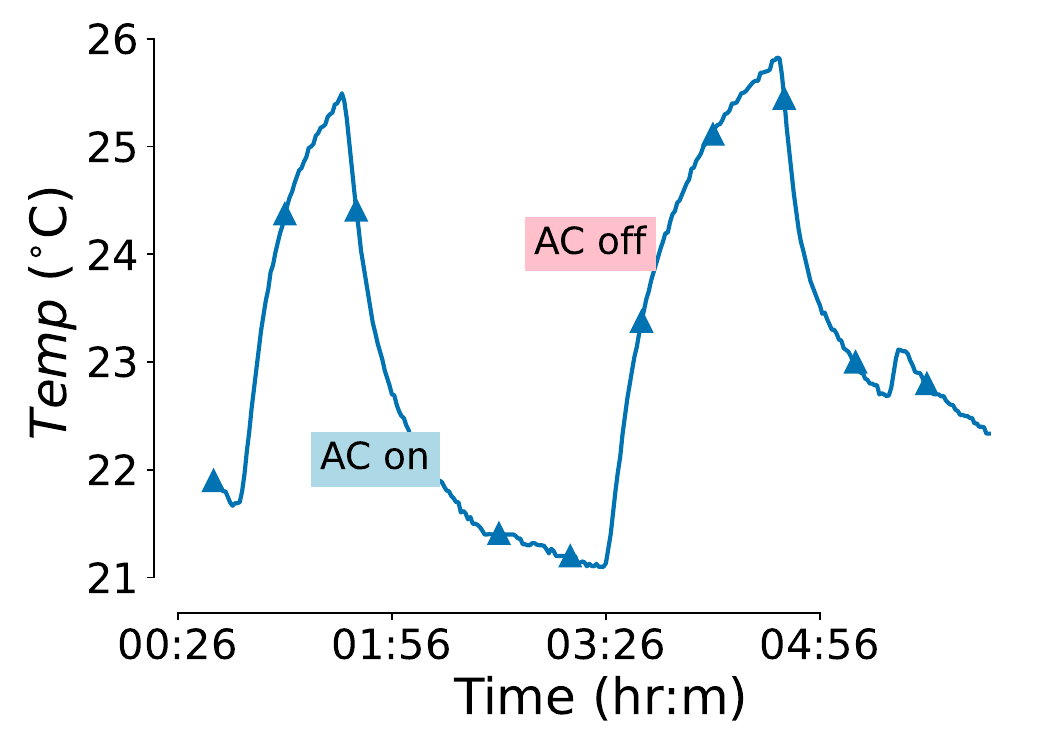}
		}
            \subfloat[Eating food\label{fig:eat}]{
			\includegraphics[width=0.25\textwidth,keepaspectratio]{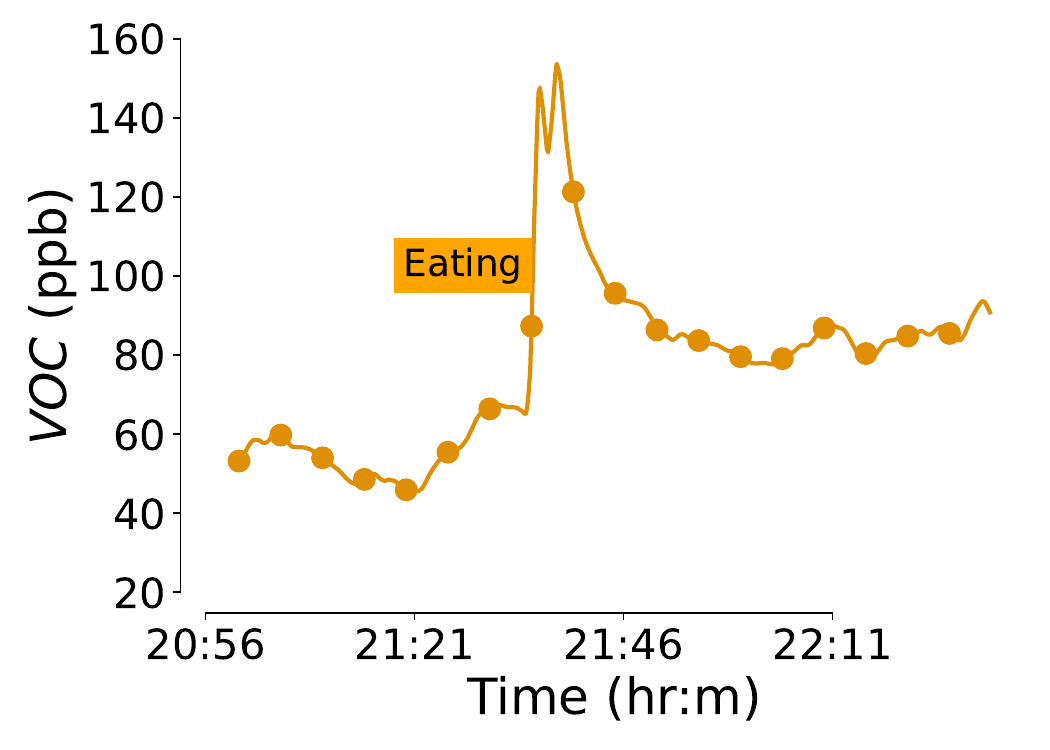}
		}
	\caption{Air monitor's measurements due to different activities - (a) Accumulation of CO\textsubscript{2} when students enter the classroom and drop upon exit during an exam, (b) Temperature change with AC on/off, (c) \changed{VOC spike when eating food.}}
	\label{fig:act}
\end{figure*}

\section{Pilot Study}
In this section, we have conducted several pilot experiments to analyze the influence of indoor activities, \changed{primarily large indoor gatherings, air-conditioning, eating, and occupancy patterns} over the measurements of the air quality monitoring devices. The observations are as follows:

\subsection{Indoor Gatherings}
To measure the impact of large indoor gatherings in indoor spaces, we collected data from a classroom during mid-semester exams at the university. Before conducting the experiments, we surveyed several classrooms and identified an ideal one equipped with split AC, and all the windows are therefore closed. The exam was scheduled for two hours, from 9:00 am to  11:00 am in the morning. In the early morning hours before the exam, we ensured that the CO\textsubscript{2} concentration was at the expected level (close to $400$ ppm) to understand the pollution footprint of 40 students gathering in the classroom. As the students started arriving at the venue at 8:45 am, the CO\textsubscript{2} accumulated as the windows were closed for effective air-conditioning. Notably, the split ACs circulate the airflow within the room, rather than pulling air from outside, to ensure effective air-conditioning with minimal energy cost~\cite{ding2023integrating}. However, this made the pollutants accumulate, which the students could not realize; instead, they felt comfortable with the cool breeze of the airflow. The pollutants reached peak levels (almost $5000$ ppm) at the last $10$ minutes of the exam. Further, we observe from the floating figure in \figurename~\ref{fig:exam} that the pollutants remain trapped in that space for a long time even after all the students have left the classroom. Therefore, pollutants due to consecutive indoor gathering can add up and result in long-term accumulation of CO\textsubscript{} in indoors.

\subsection{Air Conditioning}
The air monitors are mostly equipped with temperature and humidity sensors. Air conditioning systems directly impact the temperature. Therefore, when the AC is turned on AC, the temperature goes down and vice versa. Similar observations can be seen in \figurename~\ref{fig:ac}, where the temperature goes down from 26$^{\circ}$ C to 23 $^{\circ}$ C. The temperature starts rising as soon as the air conditioning is off. \changed{Moreover, several pollutants (i.e., CO\textsubscript{2}, VOC) also get impacted by air-conditioning based on the type of AC. We have observed that split-AC compromises ventilation to improve power efficiency, resulting in the accumulation of indoor pollutants.}

\subsection{Eating Food}
\figurename~\ref{fig:eat} shows the sudden increase in volatile organic compounds (VOC) concentration due to eating fruits near a \ourmethod{} module. We can observe elevated levels of VOC for the duration of eating activity (approx. 10 minutes). The pollutant starts normalizing as soon as the activity ends and the table is cleaned. Further, we observed that food scraps can act as a long-term pollution source until they are removed from the indoor space.

\subsection{Occupancy Patterns}
\begin{figure}
    \centering
\includegraphics[width=0.38\textwidth,keepaspectratio]{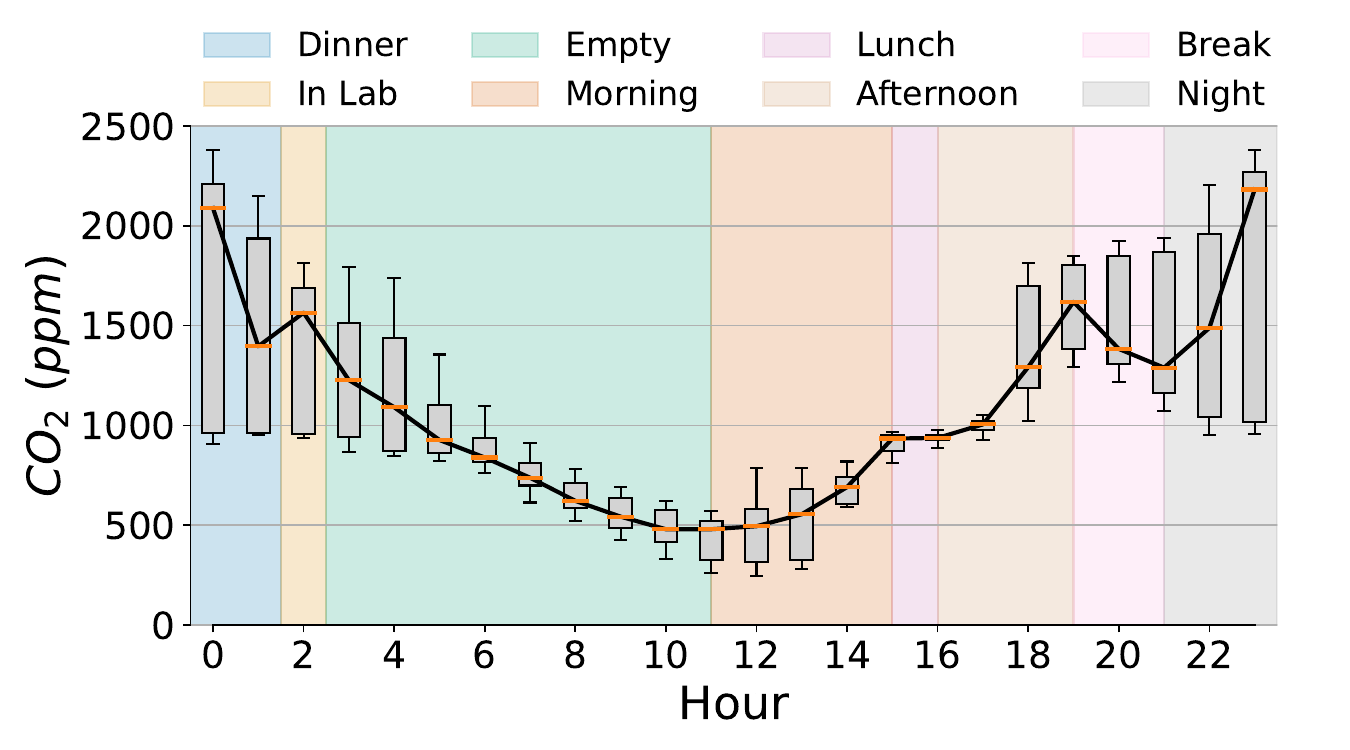}
    \caption{Variation of CO\textsubscript{2} concentration with indoor activities throughout the day. We observe that CO\textsubscript{2} correlates with indoor occupancy or gathering.}
    \label{fig:lab_co2}
\end{figure}

An overall daily pattern in CO\textsubscript{2} variation for the collected data from the research lab is shown in \figurename~\ref{fig:lab_co2}. We observe that indoor activities and occupancy influence the overall CO\textsubscript{2} levels. As shown in the figure, CO\textsubscript{2} concentration keeps rising when the lab is occupied, and during the dinner, lunch, and break hours, it falls slightly due to less occupancy. The lab members usually come to the lab at around 10 am in the morning. The CO\textsubscript{2} starts accumulating till 2:00 pm in the noon when the members go to lunch. However, the CO\textsubscript{2} remains at a similar concentration due to less ventilation. The pollutant increases during the afternoon and evening hours due to maximum occupancy before the evening break. Again, during the night hours, the CO\textsubscript{2} accumulates until the dinner break. In summary, indoor pollutants such as CO\textsubscript{2} are significantly influenced by indoor activities and occupancy patterns.

%% file: Sections/dataset.tex
\begin{figure*}
        \centering
	\captionsetup[subfigure]{}
		\subfloat[Lab Deployment\label{fig:scenario}]{
			\includegraphics[width=0.38\textwidth,keepaspectratio]{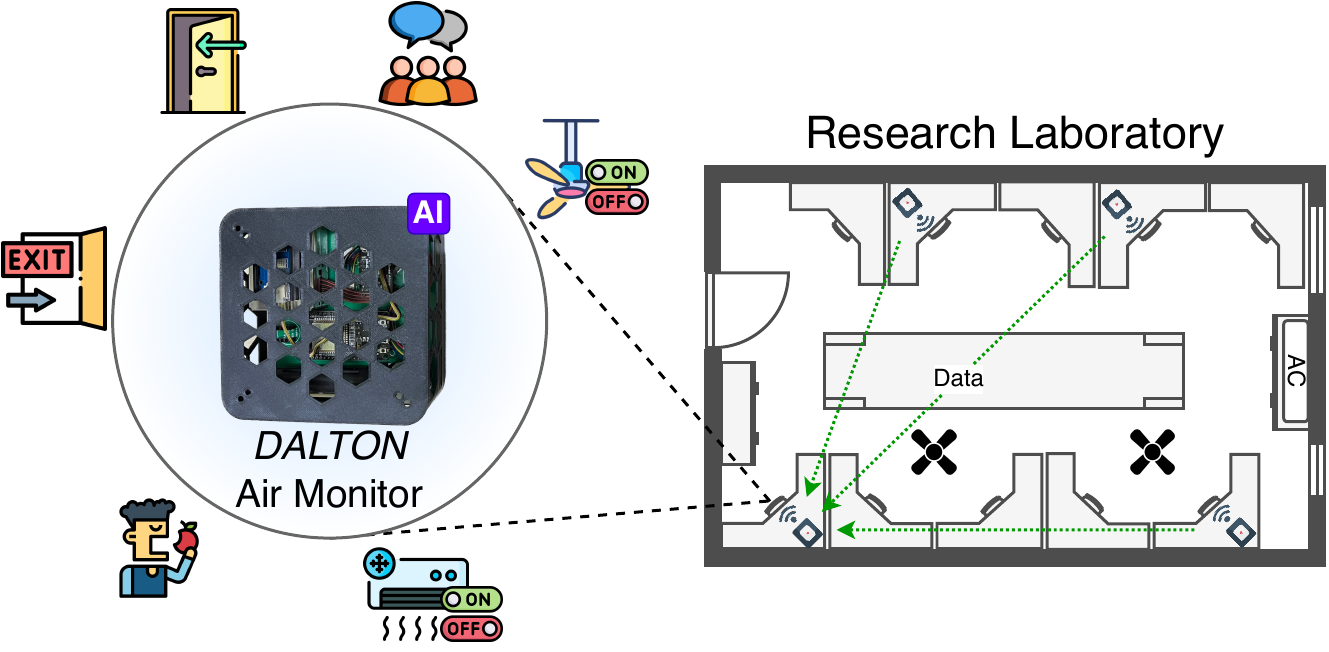}
		}
		\subfloat[Class Distribution\label{fig:data_pie}]{
			\includegraphics[width=0.38\textwidth,keepaspectratio]{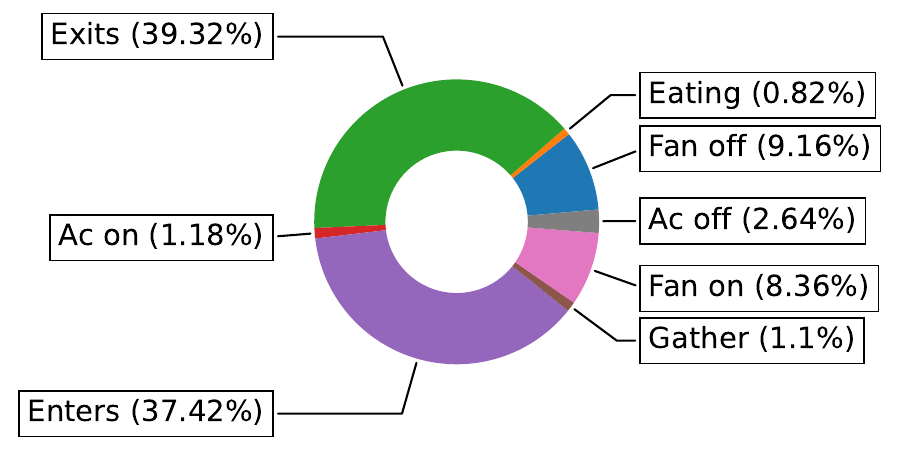}
		}
	\caption{\changed{Data collection - (a) Deployment of \ourmethod{} modules in four corners of the research Lab, (b) Proportion of the recorded activity classes in the dataset. We observe that the most frequent activities are members entering and exiting the lab, where prohibited practices like eating or gathering are very infrequent.}}
	\label{fig:dataset}
\end{figure*}

\section{Data Collection}
We have deployed the \ourmethod{} air quality monitors for three months in four corner desks of an academic research lab as shown in \figurename~\ref{fig:scenario}. In total, seven volunteers annotated their activities in the lab throughout the entire data collection process. \changed{The annotations broadly represent the engagement and attendance of each lab member (i.e., when someone exits or enters the lab), their daily behavior with electrical appliances (i.e., when the AC or Fan is turned on/off), and prohibited practices (i.e., eating and gathering in the lab). The collected dataset comprises 705 such activity annotations from the volunteers, along with changes in the pollution concentration at four strategic locations (i.e., sensors are deployed nearest to fans, AC, and sitting places to observe pollutant fluctuations). We did not instruct the volunteers to alter their daily behavior or habits, keeping their autonomy integrated with the lab space during the data collection period. \figurename~\ref{fig:data_pie} shows the class distribution of the collected dataset. From the figure, we can observe that most annotations are comprised of members entering or exiting the lab, followed by the fan on/off activity. However, AC on/off events are relatively less frequent as air-conditioning is mostly active to ensure appropriate cooling for the lab equipment and workstations. Lastly, prohibited practices like gathering and eating food in the lab are infrequent. Volunteers sometimes forget to annotate their activities when they are very busy or in a hurry. In most instances, they actively participated in the study by annotating their activities.}

\section{Feature Engineering}
\label{sec:feat}
\begin{figure*}
    \centering
    \includegraphics[width=1.0\textwidth]{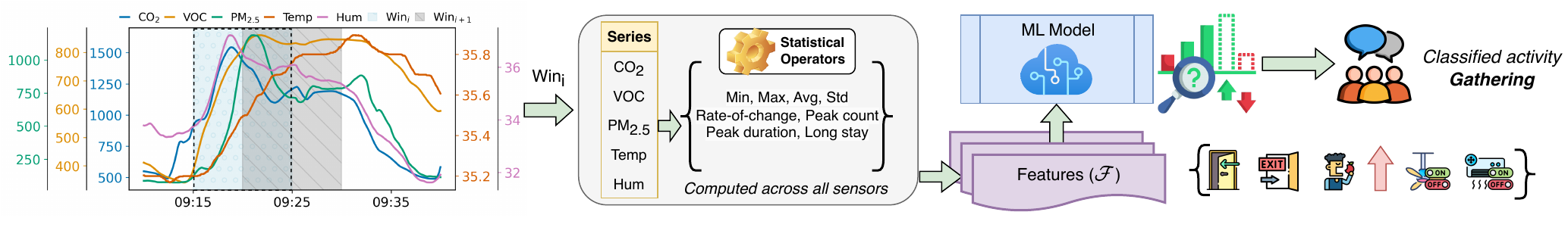}
    \caption{Overview of the data processing and modeling pipeline for activity classification. The features are computed across all neighboring devices by applying statistical operators. We use off-the-shelf ML models to classify indoor activities.}
    \label{fig:pipeline}
\end{figure*}

\changed{Based on the pollution patterns we observed during our initial pilot study and the entire data collection period, we compute a series statistical features over a sliding window of duration $\tau$\footnote{We took $\tau=10$ minutes as per empirical observations.} to capture each pollutant's abrupt changes, long and short-term accumulation, and average concentration, etc. \figurename~\ref{fig:pipeline} shows the complete set of statistical features computed for each sensor. The following are the primary reasons for selecting certain statistical operators.}

\begin{enumerate}[(I)]
    \item \textbf{Maximum and Minimum ($\mathbf{max}$, $\mathbf{min}$)}: These reflect the highest and lowest levels of pollutants recorded indoors. High maximum values indicate very poor air quality. Very low minimum values suggest good ventilation.

    \item \textbf{Standard Deviation ($\mathbf{std}$)}: This measures fluctuation for the pollutant levels. A high standard deviation indicates large swings in pollutant levels due to inconsistent ventilation or sporadic pollutant sources (e.g., gathering).

    \item \textbf{Rate of Change ($\mathbf{roc}$)}: This measures how quickly pollutant levels rise or fall. Rapid increases might occur if there's a sudden release of pollutants (e.g., eating food), while rapid decreases might indicate effective ventilation. We consider the rate of change of pollutants for both raising ($roc_{raise}$) and falling ($roc_{fall}$) edges.

    \item \textbf{Peak Count ($\mathbf{peak_c}$)}: This is the number of times pollutant levels exceed a certain unsafe threshold. Multiple peaks indicate recurring sources of pollutant-generating activities or inconsistent ventilation.

    \item \textbf{Peak Duration ($\mathbf{peak_{\Delta}}$)}: This measures the total time that pollutant levels exceeded a certain unsafe threshold. Prolonged pollution exposure resembles a long-duration activity or may point to issues with ventilation or the existence of persistent pollution sources in the indoor space.

    \item \textbf{Long Stay ($\mathbf{\Delta_{exc}}$)}: This represents the duration of moderate pollution levels above the safe threshold. Extended periods of moderate pollution indicate an overall poor indoor air quality. It also suggests inadequate ventilation or persistent sources.
\end{enumerate}
\begin{equation}
    \mathcal{F}=\bigcup_{d\in\mathcal{D}}\{f(p): \forall f,p \in \psi\times\pi \}
    \label{eq:feat}
\end{equation}
Let the set of sensing modules $\mathcal{D}=\{d_i|i=1,2,\dots N\}$, where $N$ is the number of modules, and the set of pollutants $\mathcal{\pi}=\{$$CO_2$, $VOC$, $PM_{2.5}$, $PM_{10}$, $T$, $H$$\}$. The set of computed functions over $\tau$ minute sliding window of each pollutant $\mathcal{\psi}=\{$$min$, $max$, $avg$, $std$, $roc_{raise}$, $roc_{fall}$, $peak_{c}$, $peak_{\Delta}$, $\Delta_{exc}$$\}$. Therefore, the set features, including all deployed devices, are shown in Equation~\ref{eq:feat}. \changed{As shown in the right side of \figurename~\ref{fig:pipeline}, these features are used to train simple off-the-shelf ML models to infer the activity of the lab members.}

%% file: Sections/eval.tex
\section{Evaluation}

\begin{figure}
        \centering
	\captionsetup[subfigure]{}
            \subfloat[SVM\label{fig:svm}]{
			\includegraphics[width=0.24\textwidth,keepaspectratio]{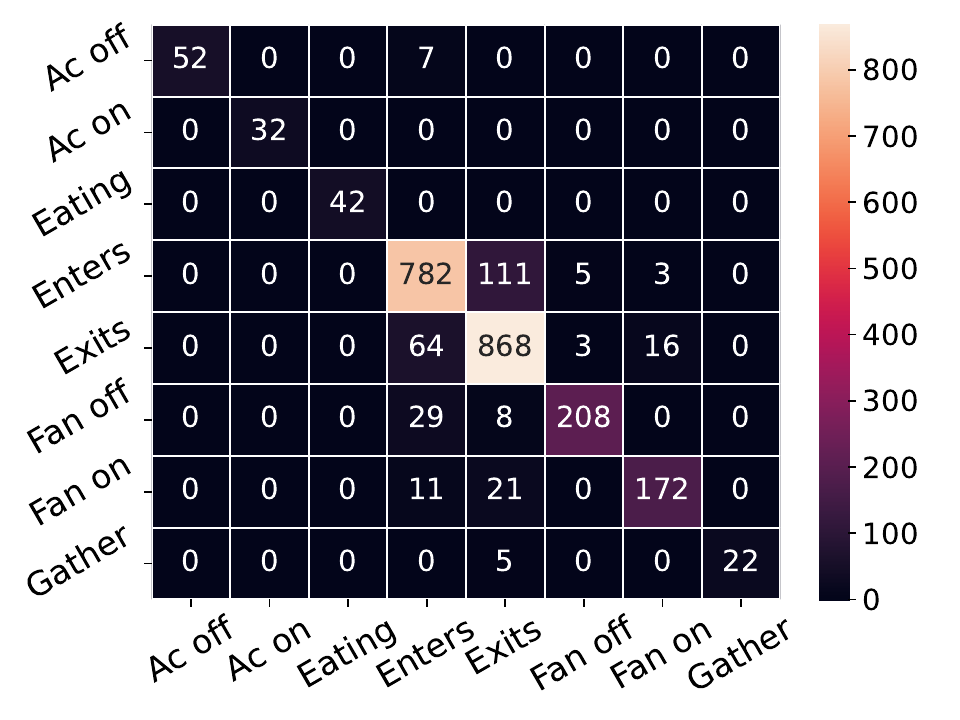}
		}
            \subfloat[Decision Tree\label{fig:dt}]{
			\includegraphics[width=0.24\textwidth,keepaspectratio]{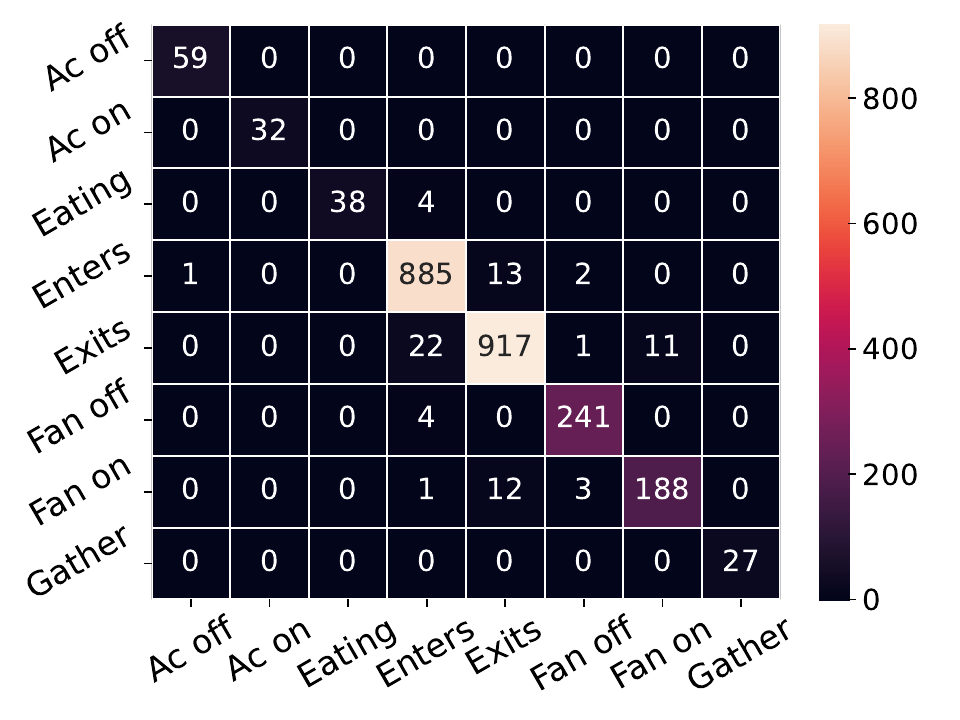}
		} \
            \subfloat[Random Forest\label{fig:rf}]{
			\includegraphics[width=0.24\textwidth,keepaspectratio]{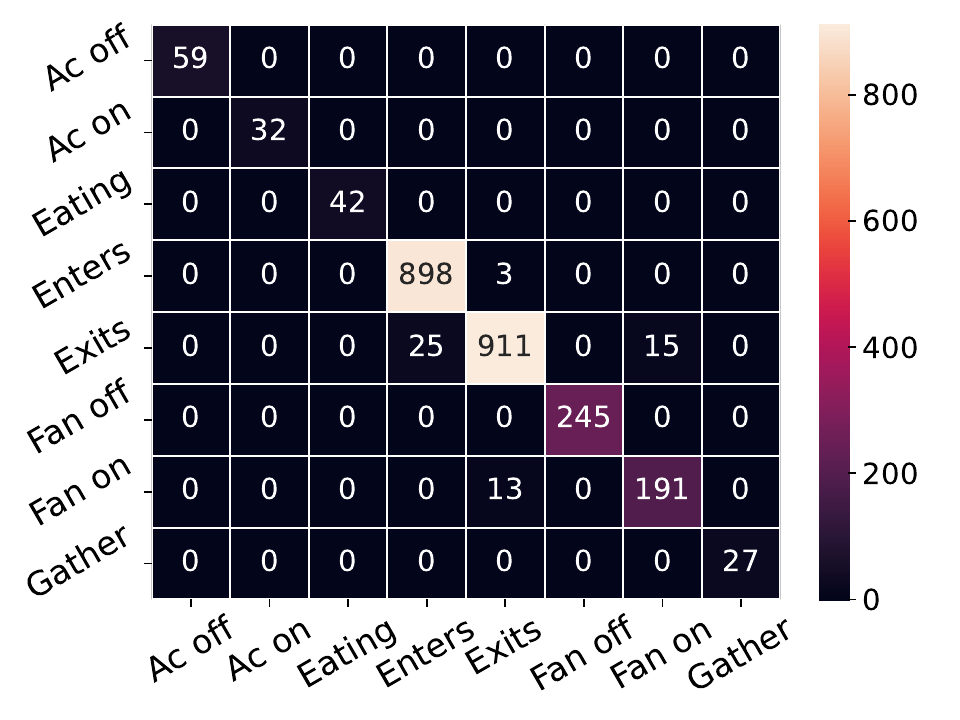}
		}
            \subfloat[Neural Network\label{fig:mlp}]{
			\includegraphics[width=0.24\textwidth,keepaspectratio]{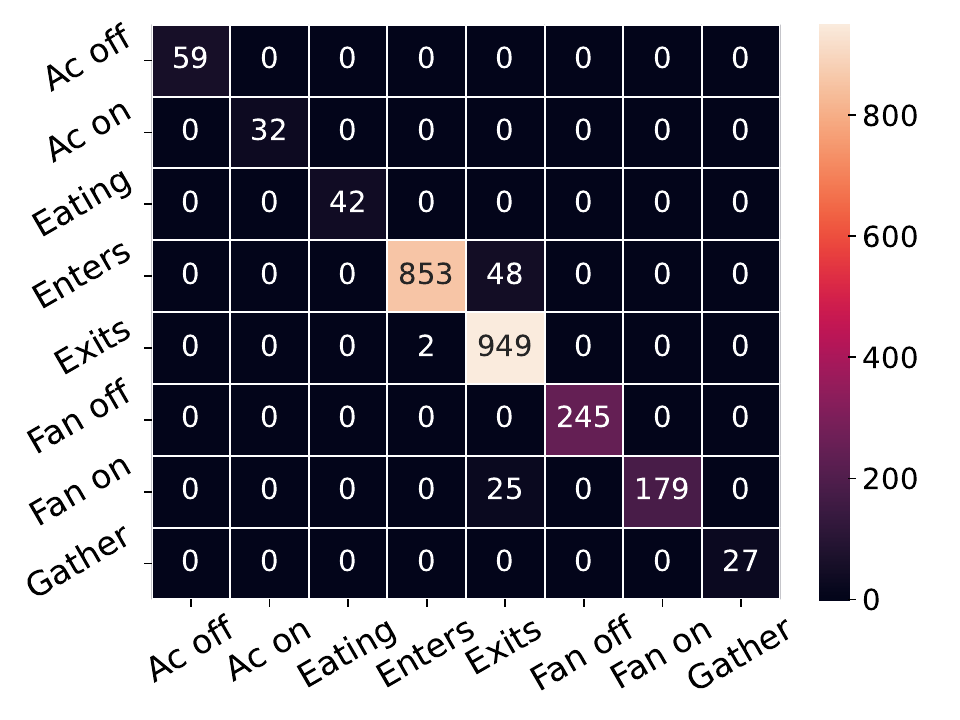}
		}
	\caption{Testing confusion matrix - (a) SVM with polynomial kernel, (b) Decision Tree with max depth 30, (c) Random Forest with max estimators 50 and depth 10, (d) Neural Network with three 64 neuron hidden layer.}
	\label{fig:conf}
\end{figure}

\begin{figure}
        \centering
	\captionsetup[subfigure]{}
            \subfloat[SVM\label{fig:svm_poly}]{
			\includegraphics[width=0.45\columnwidth,keepaspectratio]{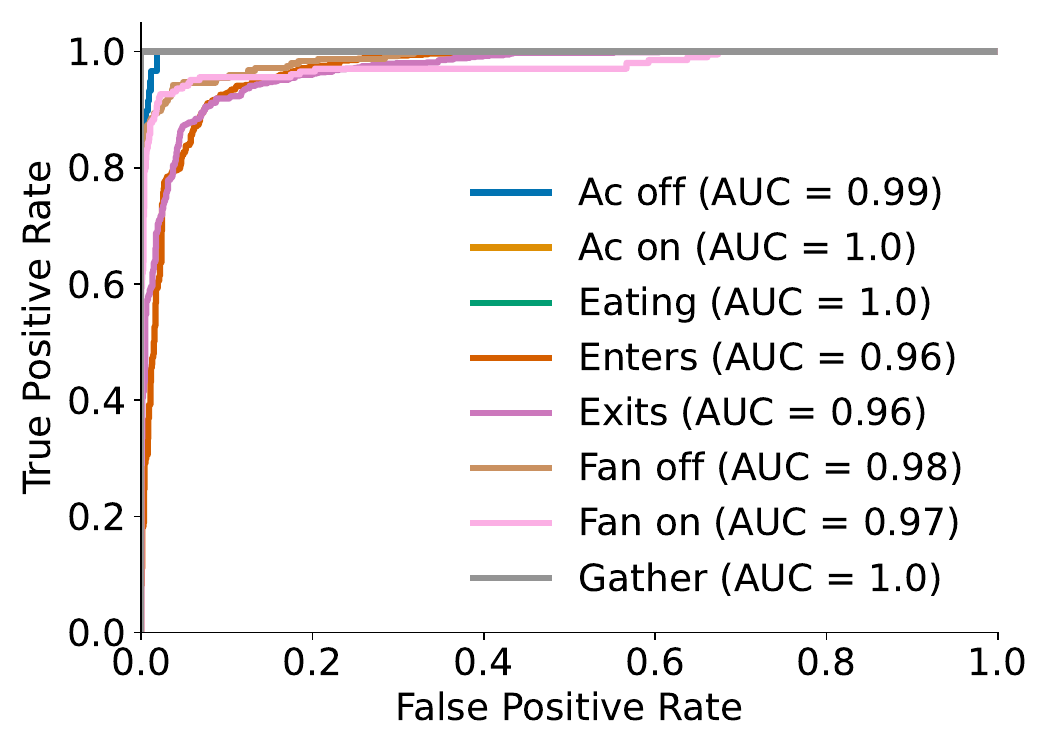}
		}
            \subfloat[Random Forest\label{fig:rf_me50}]{
			\includegraphics[width=0.45\columnwidth,keepaspectratio]{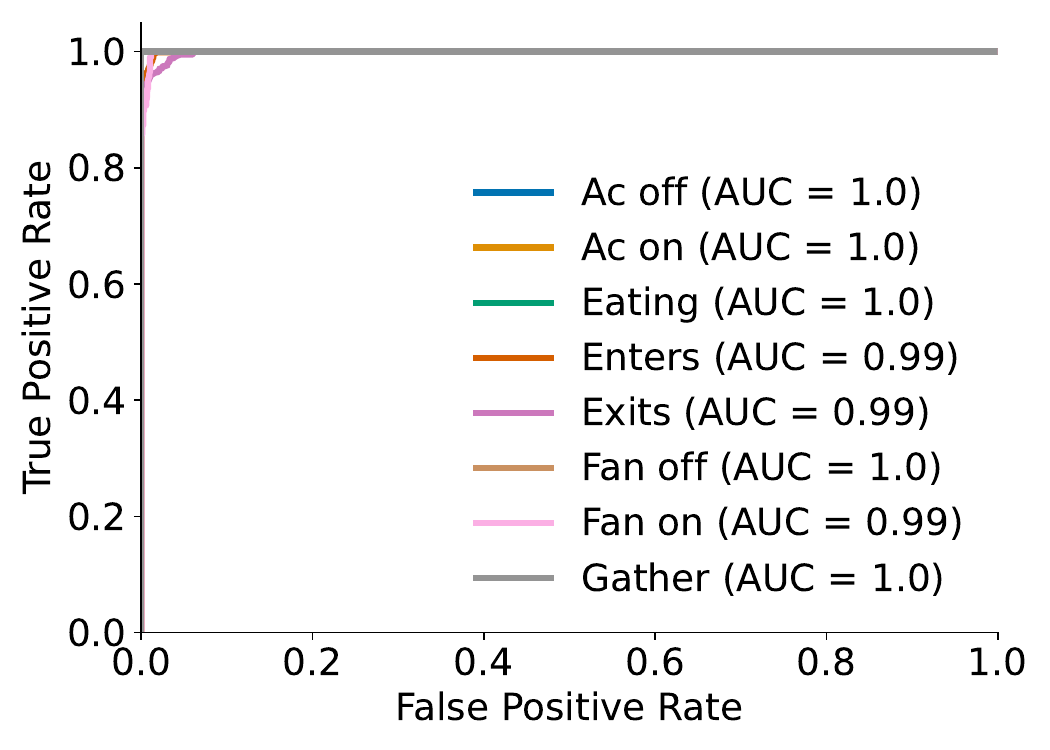}
		}
	\caption{AUC-ROC Curves - (a) SVM with polynomial kernel, (b) Random Forest with max estimators 50 and depth 10.}
	\label{fig:auc_roc}
\end{figure}

\changed{This section evaluates our setup with multiple off-the-shelf machine learning (ML) models. We have kept the models lightweight, considering the efficiency of the system. \figurename~\ref{fig:conf} shows confusion matrices of four models that performs with above 80\% F1-score during our testing. The support vector machine (SVM) shows 87.9\% F1-score with polynomial kernel. The best-performing model, random forest, shows 97.7\% testing F1-score. The respective AUC-ROC curves are shown in \figurename~\ref{fig:auc_roc}. From \figurename~\ref{fig:conf}, we observe that the models misclassify when someone enters or exits the lab. Moreover, exiting and turning on the fan is also confusing for the models. The degree of such confusion increases as we go for simpler models. For instance, random forest shows minimal confusion among these pairs as per \figurename~\ref{fig:rf}. In contrast, SVM with a polynomial kernel shows greater confusion as per  \figurename~\ref{fig:svm}. Note that the neural network with three 64-neuron hidden layers performs poorer than random forest for some instances (e.g., enter, exit) due to the cardinality and sample size of the dataset. As we evaluate for even simpler models like logistic regression and naive bayes, the overall performance significantly degrades to 76.4\% and 39.9\% F1-score, respectively.
The primary reason is interleaved activities performed by the lab members, where the pollution signatures get convoluted due to multiple factors. For instance, a member can enter the lab and turn on the fan, resulting in a mixed influence on the pollution data. Moreover, the lab protocol also plays a role by insisting that the members turn on the fans when they leave the lab. Thus, overlapped activities generate convoluted pollution signatures, making the classification task nonlinear and intractable for simpler models.}

\changed{\tablename~\ref{tab:res} summarizes the detailed evaluation of 70-30 random split and 5-fold cross-validation experiments across seven machine learning models with varying parameters. According to the table, the four-layer (three 64-neuron hidden) neural network and random forest (50 estimators of maximum depth 10) perform comparably in the random split experiments. However, the random forest shows the highest accuracy in the 5-fold cross-validation. Moreover, the random forest is more computationally efficient and lightweight for on-device execution. Therefore, random forest is the best-suited model for our dataset. The neural network suffers from high cardinality and limited sample size and may improve with a larger dataset. The best model parameters are shown in bold font in the table. Most ML models reported in \tablename~\ref{tab:res} achieve greater than 80\% testing F1-score, indicating potential applications of air quality data in monitoring restricted or private indoor spaces.}

\begin{table*}
\centering
\caption{Performance of the ML models in 70-30 random split and 5-fold cross-validation experiments in the collected dataset.}
\label{tab:res}
\resizebox{\textwidth}{!}{%
\begin{tabular}{|l|l|c|c|c|c|c|c|c|c|c|c|} 
\hline
\multirow{3}{*}{\textbf{Model}}      & \multirow{3}{*}{\textbf{Parameters}}                                    & \multicolumn{6}{c|}{\textbf{70-30 Random Split}}                                                                    & \multicolumn{4}{c|}{\textbf{5-Fold Cross-validation}}                                                \\ 
\cline{3-12}
                                     &                                                                         & \multicolumn{3}{c|}{\textbf{Training (Weighted)}}        & \multicolumn{3}{c|}{\textbf{Testing (Weighted)}}         & \multicolumn{2}{c|}{\textbf{Accuracy (Mean)}} & \multicolumn{2}{c|}{\textbf{Accuracy (Std)}}  \\ 
\cline{3-12}
                                     &                                                                         & \textbf{F1-score} & \textbf{Precision} & \textbf{Recall} & \textbf{F1-score} & \textbf{Precision} & \textbf{Recall} & \textbf{Train} & \textbf{Test}                & \textbf{Train}  & \textbf{Test}               \\ 
\hline
\multirow{3}{*}{SVM}                 & Linear kernel                                                           & 0.831             & 0.837              & 0.833           & 0.817             & 0.827              & 0.821           & 0.495          & 0.497                        & 0.0156          & 0.0176                      \\ 
\cline{2-12}
                                     & Polynomial kernel                                                       & \textbf{0.892}    & \textbf{0.894}     & \textbf{0.892}  & \textbf{0.879}    & \textbf{0.881}     & \textbf{0.879}  & \textbf{0.543} & \textbf{0.542}               & \textbf{0.0041} & \textbf{0.0075}             \\ 
\cline{2-12}
                                     & RBF kernel                                                              & 0.815             & 0.836              & 0.82            & 0.798             & 0.82               & 0.804           & 0.53           & 0.53                         & 0.0036          & 0.01                        \\ 
\hline
Naive Bayes                          & Gaussian                                                                & 0.403             & 0.727              & 0.399           & 0.399             & 0.717              & 0.39            & 0.424          & 0.419                        & 0.0105          & 0.0047                      \\ 
\hline
\multirow{4}{*}{Decision Tree}       & Max depth10                                                             & 0.949             & 0.95               & 0.949           & 0.924             & 0.924              & 0.924           & 0.967          & 0.951                        & 0.0025          & 0.0051                      \\ 
\cline{2-12}
                                     & Max depth 20                                                            & 0.992             & 0.992              & 0.992           & 0.975             & 0.975              & 0.976           & 0.992          & 0.974                        & 0.0008          & 0.0039                      \\ 
\cline{2-12}
                                     & Max depth 30                                                            & \textbf{0.992}    & \textbf{0.992}     & \textbf{0.992}  & \textbf{0.976}    & \textbf{0.976}     & \textbf{0.976}  & \textbf{0.992} & \textbf{0.975}               & \textbf{0.0007} & \textbf{0.0039}             \\ 
\cline{2-12}
                                     & Max depth 40                                                            & 0.992             & 0.992              & 0.992           & 0.976             & 0.976              & 0.976           & 0.992          & 0.975                        & 0.0007          & 0.0039                      \\ 
\hline
\multirow{4}{*}{k-Nearest Neighbour} & Neighbour 10                                                            & \textbf{0.981}    & \textbf{0.981}     & \textbf{0.981}  & \textbf{0.975}    & \textbf{0.975}     & \textbf{0.975}  & \textbf{0.985} & \textbf{0.979}               & \textbf{0.0008} & \textbf{0.0022}             \\ 
\cline{2-12}
                                     & Neighbour 20                                                            & 0.972             & 0.972              & 0.972           & 0.967             & 0.967              & 0.967           & 0.983          & 0.979                        & 0.0002          & 0.0024                      \\ 
\cline{2-12}
                                     & Neighbour 30                                                            & 0.959             & 0.96               & 0.96            & 0.956             & 0.956              & 0.957           & 0.979          & 0.976                        & 0.0015          & 0.0044                      \\ 
\cline{2-12}
                                     & Neighbour 40                                                            & 0.946             & 0.946              & 0.947           & 0.947             & 0.947              & 0.947           & 0.975          & 0.972                        & 0.0021          & 0.0036                      \\ 
\hline
Logistic Regression                  & --                                                                      & 0.791             & 0.801              & 0.796           & 0.764             & 0.777              & 0.77            & 0.581          & 0.577                        & 0.0088          & 0.0145                      \\ 
\hline
\multirow{3}{*}{Random Forest}       & \begin{tabular}[c]{@{}l@{}}Max estimator 30\\Max depth 10\end{tabular}  & 0.988             & 0.988              & 0.988           & 0.979             & 0.979              & 0.979           & 0.989          & 0.977                        & 0.0005          & 0.0021                      \\ 
\cline{2-12}
                                     & \begin{tabular}[c]{@{}l@{}}Max estimator 50\\Max depth 10\end{tabular}  & \textbf{0.988}    & \textbf{0.988}     & \textbf{0.988}  & \textbf{0.977}    & \textbf{0.977}     & \textbf{0.977}  & \textbf{0.989} & \textbf{0.979}               & \textbf{0.0006} & \textbf{0.0049}             \\ 
\cline{2-12}
                                     & \begin{tabular}[c]{@{}l@{}}Max estimator 100\\Max depth 10\end{tabular} & 0.99              & 0.99               & 0.99            & 0.979             & 0.979              & 0.979           & 0.989          & 0.977                        & 0.0012          & 0.0037                      \\ 
\hline
\multirow{4}{*}{Neural Network}      & Hidden [64, 64]                                                         & 0.981             & 0.981              & 0.981           & 0.973             & 0.973              & 0.973           & 0.925          & 0.92                         & 0.0229          & 0.0165                      \\ 
\cline{2-12}
                                     & Hidden [64, 64, 64]                                                     & \textbf{0.982}    & \textbf{0.983}     & \textbf{0.982}  & \textbf{0.978}    & \textbf{0.979}     & \textbf{0.978}  & \textbf{0.947} & \textbf{0.943}               & \textbf{0.0104} & \textbf{0.0135}             \\ 
\cline{2-12}
                                     & Hidden [128, 128]                                                       & 0.981             & 0.982              & 0.981           & 0.979             & 0.979              & 0.979           & 0.912          & 0.91                         & 0.0116          & 0.0114                      \\ 
\cline{2-12}
                                     & Hidden [128, 128, 128]                                                  & 0.982             & 0.982              & 0.982           & 0.978             & 0.978              & 0.978           & 0.95           & 0.946                        & 0.0174          & 0.0203                      \\
\hline
\end{tabular}}
\end{table*}

%% file: Sections/conclusion.tex
\section{conclusion}
This paper explores potential side-channel applications of pollutant measurements from ubiquitous air monitoring solutions to identify indoor activities. Therefore, sharing pollution data with third parties may cause privacy concerns, as with such capabilities, one can carry out indoor surveillance without the end user's consent. In this work, we have collected pollution data annotated with eight indoor activities (i.e., enter, exit, fan on, fan off, AC on, AC off, gathering, eating) in a research lab over three months. Our analysis highlights that indoor pollutants are greatly influenced by the activities being performed. We can predict the underlying activity from the pollution data with 97.7\% F1-score using a simple light-weight random forest model. \changed{In the future, we plan to increase the set of detectable activities by developing a generalized framework transferable to restricted (e.g., research lab) and private (i.e., residential) spaces.}

%% file: Sections/ethics.tex
\changed{\section{Ethical Clearance}
The institute's ethical review committee has reviewed and approved the field study (Order No: IIT/SRIC/DEAN/2023, dated 31 July 2023). In addition, all participants signed consent forms in order for pollutant measurements and activity annotations to be used for non-commercial research. Further, we have ensured privacy and autonomy of the participants during data collection and provided information required to encourage future research on the applications of indoor air pollution.}

%% file: Sections/acknow.tex
\changed{\section{acknowledgement}
The authors would like to thank the anonymous reviewers for the constructive comments, which have helped to improve the overall presentation of the paper. The research of the first author is supported by the Prime Minister Research Fellowship (PMRF) in India through grant number IIT/Acad/PMRF/SPRING/2022-23, dated 24 March 2023. The work is also supported by AI4ICPS IIT Kharagpur Technology Innovation Hub award number(s): TRP3RD3223323 Dated 22 Feb 2024, and Google's Award for Inclusion Research 2023 for the project proposal ``\textit{AI-Assisted Distributed Collaborative Indoor Pollution Meters: A Case Study, Requirement Analysis, and Low-Cost Healthy Home Solution For Indian Slums.''}.}